\documentclass[aps,pre,twocolumn,superscriptaddress]{revtex4}

\usepackage{graphicx}

\begin{document}

\title{\textbf{Passive Retrieval of Rayleigh Waves in Disordered Elastic Media}}


\author{Eric LAROSE}
\email[]{eric.larose@ujf-grenoble.fr}

\affiliation{Laboratoire de G\'eophysique Interne et Tectonophysique (LGIT) Universit\'e Joseph Fourier, CNRS UMR 5559, Grenoble, France}
\affiliation{Laboratoire Ondes Acoustiques (LOA), Universit\'e Paris 7, CNRS UMR 7587, ESPCI, Paris, France }

\author{Arnaud DERODE}
\affiliation{Laboratoire Ondes Acoustiques (LOA), Universit\'e Paris 7, CNRS UMR 7587, ESPCI, Paris, France }

\author{Dominique CLORENNEC}
\affiliation{Laboratoire Ondes Acoustiques (LOA), Universit\'e Paris 7, CNRS UMR 7587, ESPCI, Paris, France }

\author{Ludovic MARGERIN}
\affiliation{Laboratoire de G\'eophysique Interne et Tectonophysique (LGIT) Universit\'e Joseph Fourier, CNRS UMR 5559, Grenoble, France}

\author{Michel CAMPILLO}
\affiliation{Laboratoire de G\'eophysique Interne et Tectonophysique (LGIT) Universit\'e Joseph Fourier, CNRS UMR 5559, Grenoble, France}

\date{\today}

\begin{abstract}
When averaged over sources or disorder, cross-correlation of diffuse fields yield the Green's function between two passive sensors. This technique is applied to elastic ultrasonic waves in an open scattering slab mimicking seismic waves in the Earth's crust. It appears that the Rayleigh wave reconstruction depends on the scattering properties of the elastic slab. Special attention is paid to the specific role of bulk to Rayleigh wave coupling, which may result in unexpected phenomena like a persistent time-asymmetry in the diffuse regime. 
\end{abstract}


\maketitle

\section{INTRODUCTION}
Whatever the type of waves involved, knowing the Green's function of an heterogeneous medium is the key to many essential applications like imaging, communication or detection. In the last twenty years or so, mesoscopic physics has intensely studied wave phenomena in strongly disordered media: weak and strong localization, radiative transfer and diffusion approximation etc. \cite{vanrossum1999}.  But the \textit{exact} Green's function of a complex medium is not easily tractable, and usually theoreticians study ensemble-averaged quantities like statistical correlations of intensities or wave fields. Moreover, from an experimental point of view, it is not always possible to measure the Green's functions of a complex medium because it requires controllable arrays of sources and receivers that do not perturb the medium. In the context of laboratory ultrasound, this is (nearly) routine; but in other fields of wave physics like seismology, distant sources are not controllable. In that respect, a good deal of publications followed recent works by Weaver. He proposed to cross-correlate the diffuse wave fields obtained at two passive sensors and showed it yields the elastic Green's function between the two receivers, as if one of the receivers was a source \cite{weaver2001b,lobkis2001}. That correlations performed on passive sensors should yield the wave travel times is not that new. This principle was applied to helioseismology in the 90's where it provided tomographic images of the Sun's interior \cite{duvall1993,rickett2000}. Beyond travel time reconstruction, Weaver's experimental retrieval of 
exact Green's functions was a real breakthrough.\\

Weaver's work has been followed by different contributions. Some theoretical works are based on the ergodic approximation where time and ensemble average coincide. This approach is very useful for estimating the role of scattering in the correlation asymmetry \cite{vantiggelen2003,malcolm2004}. When several sources are available, another possibility is to average correlations over sources without moving the receivers. This was done in underwater acoustics \cite{roux2004}. Sabra \textsl{et al}~\cite{sabra2005} showed the possibility of recovering the entire Green's function of the sea waveguide (especially the late contributions of multiples) and proposed a model for estimating the signal-to-noise ratio of the correlations. Derode \textit{et al}~\cite{derode2003a} proposed to interpret the Green's function reconstruction in terms of a Time-Reversal analogy, and showed that all the benefits of time-reversal devices in multiple scattering media could be fruitfully applied to "passive imaging" from correlations \cite{derode2003b,larose2004b}. This idea is based on the mathematical principle of the representation theorem \cite{akirichards}, equivalently referred to as the Helmholtz-Kirchhoff theorem \cite{cassereau1992}: if a perfect series of receivers has been sensing the wave field for ages, then one can mathematically have access to the wave field anytime and anywhere in the area enclosed by the sensors. Very recently, this principle was fully developed and applied to elastic waves in open media by Wapenaar \cite{wapenaar2004}.\\

\begin{figure}[htbp]
\includegraphics[width=8.6cm]{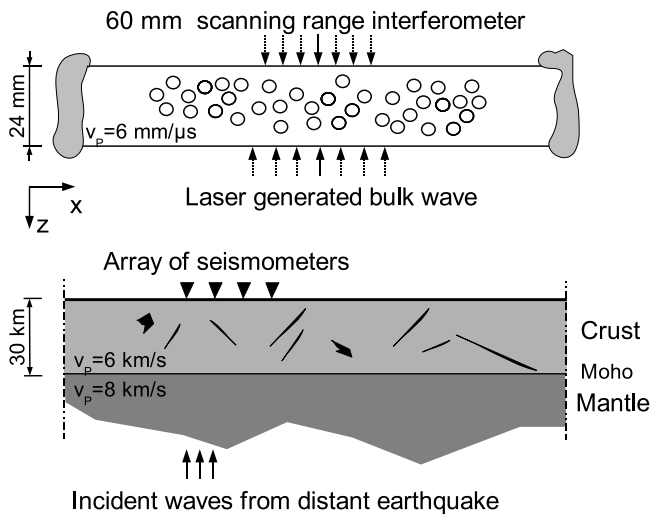}
\caption{\label{schema_analogie}$10^6 \rightarrow 1$ scaled analogy between the Earth crust and our ultrasonic experiment. In the crust, the scattering mean free path was estimated in Mexico~\cite{margerin1998} at 1~Hz: $\ell^*=30$~km. In the aluminum waveguide we numerically found $\ell^*=5.5$~mm at 1~MHz.}
\end{figure}

From the very start of Weaver's work, the correlation principle was successfully applied to seismic waves \cite{campillo2003a,shapiro2004,shapiro2005}. The most energetic part of the Green's function between two seismic stations (Rayleigh and Love wave trains) was retrieved from passive correlations of either coda waves or records of seismic noise. As we mentioned earlier, obtaining impulse responses without a controllable source is of high interest in seismology since it gives the possibility of simulating very energetic and punctual earthquakes everywhere around the Earth, and do imaging without a controllable source ("passive imaging"). Results are quite encouraging although several theoretical problems remain unsolved. They concern issues very specific to the Earth. In particular elastic sources (earthquakes) are naturally never arranged in such a way that they would perfectly surround a couple of seismic stations. In addition, for distant earthquakes, the propagation directions of incident bulk waves are mostly vertical, and in a vertically layered medium they would not couple with Rayleigh waves. So, why do we observe a Rayleigh wave train in the correlation of waves generated by bulk sources? The Green's Function retrieval is linked to equipartition, which is due to wave scattering. So, what is the influence of scattering within the Earth crust in that process?\\

Here we present experimental results obtained in the lab with laser-induced and laser-detected ultrasonic waves propagating in a heterogeneous slab mimicking the Earth crust. Lots of previous experimental articles applied the "passive imaging" technique to acoustic waves. Here we propose to investigate the emergence of the Green's function in the correlations of elastic waves propagating in a solid heterogeneous medium with a free surface. Because field experiments are tedious and natural environment are mostly unknown (especially the scattering properties), we propose to build an Earth crust model at the scale $1/10^{-6}$ (presented in fig.~\ref{schema_analogie}). Waves will be sensed at ultrasonic frequencies at the free-surface of an elastic open medium. In our experiment surface waves are not initially excited. This is different from the work of Malcolm \textsl{et al}~\cite{malcolm2004}, where ultrasonic Rayleigh waves were generated on the same surface they were measured. In addition they used a finite cylindrical medium with possibly round trip wave trains whereas our experiment is conducted in a nearly open medium. In our configuration, we would not expect the Rayleigh wave to be reconstructed in the correlation, except if scattering is present and mode conversion between Rayleigh and bulk waves occurs. To verify this assumption and study the role of mode conversion, we used two samples of identical dimensions, one with scatterers the other without.\\

The next section describes the experimental setup and the propagation medium. Section III presents a short theoretical study of the scattering properties (scattering-cross sections $\sigma$ of a single scatterer and the transport mean-free paths $\ell_P^*$  and $\ell_S^*$ of the heterogeneous sample). This study is supported by experimental measurements. In section IV, the passive imaging technique is applied to records acquired at the free surface. Time and frequency analysis are proposed, and a brief discussion on the time-symmetry of the correlations concludes the article.

\section{EXPERIMENTAL SET-UP}

\begin{figure}[htbp]
\includegraphics[width=8.6cm]{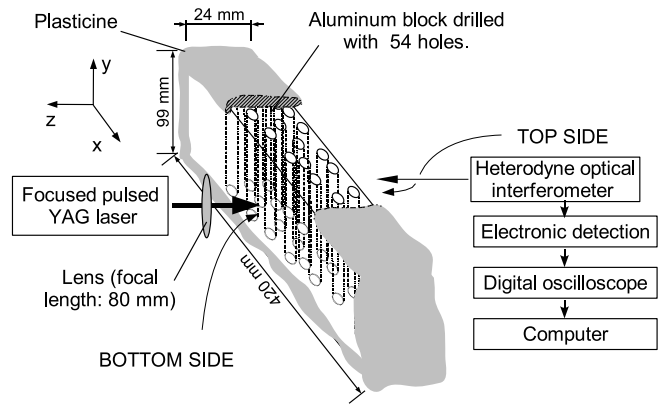}
\caption{\label{schema_manip}Experimental setup : The propagation medium is an aluminum block drilled with 54 vertical cylindrical holes (diameter: 4~mm). A  Q-switched Nd:YAG
laser shoots on the bottom side  100 pulses (24~ns duration and 280~MW/cm$^2$ intensity each).  On the top side a heterodyne interferometer senses the vertical displacement of the aluminum/air interface. This measure is repeated at 7 different locations ($X_0=0$~mm$\rightarrow X_6=60$~mm) for each source position. Plasticine was stuck on the edges and the four lateral sides to  mimic absorbing boundary conditions and avoid the generation of Rayleigh waves by mode conversion at the edges.}
\end{figure}

The experimental setup is depicted in fig.~\ref{schema_manip}. It was designed to mimic the propagation and scattering of elastic waves through the Earth crust, at ultrasonic frequencies (0.8-3.2~MHz). A duraluminum slab whose dimensions are roughly $10^{-6}$ those of the crust was used. Fifty-four cylindrical holes (radius $a = 2$~mm) were drilled at random along direction $y$ so that the waves propagating through the slab could undergo multiple scattering. The 2-D spatial Fourier transform of the hole positions was calculated and is almost perfectly flat in the range of ultrasonic wavelengths involved in the experiment, which confirms the absence of spatial correlation between the holes. The density of scatterers was $n=0.0105~$mm$^{-2}$. Ideally, the slab should have had infinite dimensions along $x$ and $y$. To approach this condition, we stuck a thick layer of dense plasticine on the lateral sides and edges of the duraluminum sample. It was aimed at creating absorbing boundary conditions and avoiding the generation of Rayleigh waves by mode conversion at the edges. The energy decay time $\tau_{abs}$ was initially found to be 23,000~$\mu$s. With the plasticine it decreased to 120~$\mu$s (see next section for a detailed discussion on the absorption time). We therefore simulated an open slab in the $x$ and $y$ direction with a free surface at the top. Rigorously, the earth crust is a waveguide that partially leaks energy through the Moho to the underlying mantle. To perfectly match this feature we should have placed a infinite medium with a different impedance at the bottom side of the aluminum slab. Yet we think that our results and conclusions do not suffer too much from this omission: the central point of our set-up is to mimic an elastic and scattering medium, infinite in the horizontal directions with a free surface on the top.\\

\begin{figure}[htbp]
\includegraphics[width=5cm]{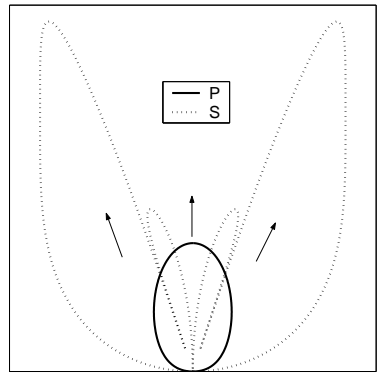}
\caption{\label{directivite}Directivity pattern (linear scale) of the laser source. Rayleigh waves are not taken into account since they are absorbed at the edges. The experiments were carried out in the ablation regime, where bulk waves are much more energetic than in the thermoelastic regime. Compressional wave directivity: thick line, shear wave directivity: dotted line.}
\end{figure}

The source we employed to simulate earthquakes was a Q-switched Nd:YAG laser that shot 24~ns pulses at the bottom side of the slab (each pulse energy : 9~mJ). Two regimes of elastic wave generation are possible with a laser source \cite{scruby1990}. When the surfacic intensity of the pulse is weak ($<15~$MW/cm$^2$), the surface is locally heated up and its dilation creates Rayleigh waves (thermoelastic regime). At higher intensity, the laser evaporates part of the metal. In this ablation regime, both Rayleigh and bulk waves are generated. Theoretical radiation patterns are displayed on fig.~\ref{directivite} for compressional and shear waves. Rigorously, such a source is not truly reproducible since the laser impact can damage the surface. To make the experiment as reproducible as possible while staying in the ablation regime, the shot intensity was no more than 280~MW/cm$^2$. A 1~ms record was acquired 100 times without any observable change. In the following experiments, for a satisfactory signal to noise ratio, each impulse response was averaged over 100 consecutive shots.\\

As to the detection of the free surface motion, it was achieved with a contactless and quasi-punctual device: a heterodyne optical interferometer developed by Royer \textsl{et al}~\cite{royer1986} which has the advantage of a very broadband response (20~kHz-45~MHz) and a sensitivity of {$10^{-4}$\AA$/\sqrt{Hz}$}. It was mounted perpendicularly to the slab  and then provided us with the absolute vertical component of the free surface displacement (top side), with a fine spatial resolution; the size of the laser spot was $\sim 100~\mu$m whereas the typical elastic wavelengths here are ranging between 1 and 10~mm. This is similar to seismology, where sensors are nearly punctual compared to the wavelengths considered (several kilometers at 1~Hz). However seismic sensors usually provide time records of the three components of the displacement field. Here the interferometer only measured the vertical movements of the free surface.\\ 

In an elastic body, three different kinds of wave polarization are possible. Compressional (or longitudinal) waves are analogous to acoustic waves in fluids (velocity $v_P=6.32~$mm/$\mu$s in duraluminum). Shear (transverse) waves have two possible polarizations (velocity $v_S=3.13$~mm/$\mu$s): one we call SV (Vertical) in the \textit{x-z} plane (see fig.~\ref{schema_analogie}) and one SH (Horizontal) in the \textit{x-y} plane. SH waves have no contribution in the $z$ direction and therefore will not be detected by the interferometer. In addition to bulk waves, surface waves exist but here only Rayleigh waves ($v_R=2.9$~mm/$\mu$s) will be taken into account since the others cannot be detected (no vertical displacement). The shortest wavelength in the aluminium slab is 0.9 mm, which is much greater than the duraluminum alloy grain size. Since the orientation of the grains is random, we consider the alloy to be isotropic for elastic waves in the frequency band of interest. Scattering at the grain edge is presumably  also negligible compared to scattering by the void cylinders.\\

The overall translational symmetry along \textit{y} of both the free surface and the cylindrical scatterers avoid any coupling between SH mode and the other SV and P modes. Therefore SH waves will not be considered in our article and SV waves will be referred to as S (shear) waves. The wave propagation in our experiment will be treated as 2-dimensional (and quasi 1-D for the surface Rayleigh waves). Waves initially propagating in the $y$ direction are rapidly absorbed by the plasticine and lost.\\

\begin{figure}[htbp]
\includegraphics[width=8.6cm]{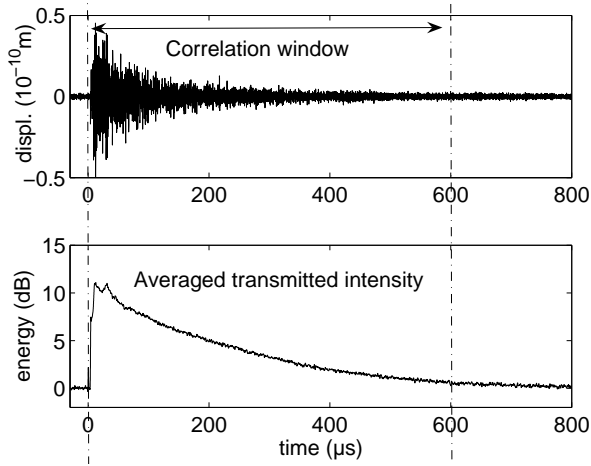}
\caption{\label{signaux}Top: typical waveform obtained through the scattering slab around 1~MHz. The sensitivity of the heterodyne interferometer is $10^{-4}$\AA$/\sqrt{Hz}$ corresponding the optimal level of optical reflection on the sensed surface. After averaging over 100 records we reached the precision of $10^{-2}~$\AA. Bottom: intensity averaged over several source/sensor positions.}
\end{figure}

The laser source and the laser interferometer could be translated independently: $35$ sources  and 7 sensing positions were used during the experiment, providing us with a set of $35\times 7$ impulse responses. A typical waveform is depicted on fig. \ref{signaux}. Around 1~MHz it is lasting nearly 800~$\mu$s, and shows a long diffusive decay comparable to the seismic coda. Due to the strong scattering on the cylindrical cavities, no top-bottom reflection was observed in the data. This confirms the high diffusive nature of the propagation in the scattering slab. The relevant scattering properties are discussed and evaluated in the next section.\\

\section{Wave scattering and transport properties}
\subsection{Scattering cross-section of an empty cylinder}

In order to evaluate the amount of scattering and mode conversion, we calculated the differential cross-section ${\partial \sigma } \over {\partial \theta}$ and the total scattering cross-section $\sigma$ of a cylindrical void in a elastic medium excited by a compressional or shear plane wave. A brief description of the calculation is given in the Appendix. For a detailed derivation we refer to \cite{paomow1973,faran1951,liu2000}. The differential scattering cross-section gives the angular distribution of the scattered surfacic intensity, normalized by the incident surfacic intensity. The total elastic cross-section is $\sigma=\int\frac{\partial \sigma }{\partial \theta} d\theta$. In 2D it has the dimensions of a length. It corresponds to the scattering strength of an object at a given frequency. In an elastic medium, mode conversion can occur and different cross-sections must be considered. In the case of an incident compressional wave, they are noted $\sigma_{PP}$, $\sigma_{PS}$ and $\sigma_P=\sigma_{PP}+\sigma_{PS}$, respectively for the P to P, P to S and total P elastic cross-sections. We also calculated the elastic cross-sections for an incident shear wave (S), $\sigma_{SP}$, $\sigma_{SS}$ and $\sigma_S=\sigma_{SP}+\sigma_{SS}$.

\begin{figure}[htbp]
\includegraphics[width=8.6cm]{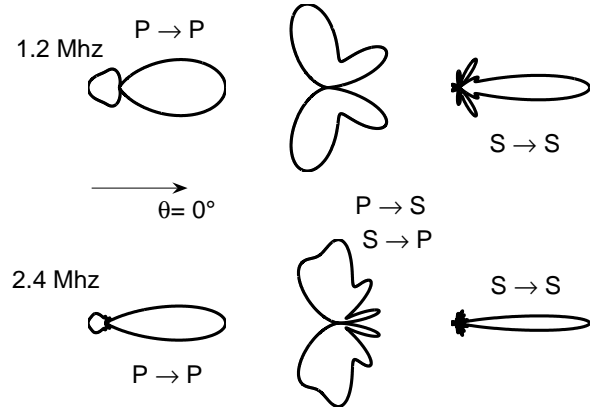}
\caption{\label{ScatDiffCrossSec1_2_3MHz} Differential scattering cross-sections of a cylindrical cavity calculated for a compressional (P) or shear (S) incident plane wave around 1.2~MHz (top) and 2.4~MHz (bottom). The scattered wave is either compressional (P) or transverse (S). Each pattern is normalized by its maximum.}
\end{figure}

\begin{figure}[htbp]
\includegraphics[width=8cm]{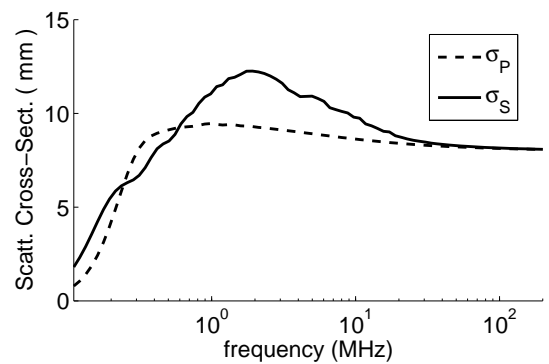}
\caption{\label{totalcrosssection} Elastic ($\sigma$)  scattering cross-sections calculated for shear (S) and compressional (P) plane wave impinging on a cylindrical void with diameter 4~mm. Between 0.8 and 3.2~MHz, scattering is stronger for shear waves. At high frequencies the elastic cross-sections tend to the limit of twice the geometrical diameter. }
\end{figure}

The differential cross-sections plotted on fig.~\ref{ScatDiffCrossSec1_2_3MHz} have been computed at 1.2~MHz and 2.4~MHz frequencies. The elastic scattering sections are plotted in fig.~\ref{totalcrosssection} for frequencies ranging from 0.1~MHz to 200~MHz. 
In average in the frequency band of interest (0.8-3.2~MHz), we obtained $\sigma_P=9.2$~mm. This value is comparable to measurements by White~\cite{white1958}. The same calculations were conducted for an incident shear wave, we found an average of $\sigma_S=12$~mm in the 0.8-3.2~MHz frequency band.\\

\subsection{Transport properties}

When the elastic wave propagates through the aluminum slab drilled with holes, it undergoes multiple scattering. Let $\varphi(t)$ (resp. $\varphi_0(t)$) be the vertical displacements sensed at the free surface through the scattering (resp. homogeneous) medium. Classically, this field is split into two contributions: the coherent and the incoherent part.
The coherent wave is the ensemble-averaged field $\left\langle \varphi(t)\right\rangle$ (averaged over disorder configurations, here: cylinder's positions). We underline the difference between the coherent and the ballistic wave (i.e. the first arrival). For a detailed discussion about scattering effect on coherent and ballistic waves, see \cite  {derode2001a}. Away from resonances, the coherent wave can be roughly thought of as an attenuated version of the direct wavefront $\varphi_0(t)$. When there is no intrinsic dissipation, the energy of the coherent wave decays with the slab thickness $H$ as $e^{-H/\ell}$, where $\ell$ is the elastic mean free path. Assuming a dilute set of scatterers, the elastic mean-free path is simply related to the scatterers density $n$ and their elastic cross-section $\sigma$: 
$$
\ell=\frac{1}{n\sigma}
$$
From the theoretical scattering cross section calculated above, we find $\ell_P = 10.3$~mm. In order to measure the mean-free path experimentally, we used two aluminum slabs of exactly the same dimensions. The first served as a reference and provided measurements of $\varphi_0(t)$ for different source-sensors positions. The second one was drilled with holes. By translating the source-receiver device along the slab, we achieved something very similar to a configurational averaging and measured the energy of the coherent wave $\left\langle \varphi(t)\right\rangle^2$. Between 0.8 and 3.2~MHz, we obtained $\ell_P=9\pm2.5$~mm from these experiments.

The intensity of the incoherent part was also studied. The time evolution of the averaged incoherent intensity $I(t)=\left\langle \varphi(t)^2\right\rangle$ is governed by another parameter: the transport mean free path $\ell^*$. In an elastic body, transport quantities have been theoretically defined by~\cite{turner1998}:
\begin{equation}
\ell^*_P=\frac{1}{n}\frac{\sigma_S-\sigma^*_{SS}+\sigma^*_{PS}}{(\sigma_P-\sigma^*_{PP})(\sigma_S-\sigma^*_{SS})-\sigma^*_{PS}\sigma^*_{SP}}
\end{equation}

\begin{equation}
\ell^*_S=\frac{1}{n}\frac{\sigma_P-\sigma^*_{PP}+\sigma^*_{SP}}{(\sigma_P-\sigma^*_{PP})(\sigma_S-\sigma^*_{SS})-\sigma^*_{PS}\sigma^*_{SP}}
\end{equation}
with $\sigma^*=\int\frac{\partial \sigma }{\partial \theta}cos(\theta) d\theta$. It was evaluated numerically: $\ell^*_P\approx \ell^*_S=5.5$~mm. In an experiment, this parameter is very hard to measure with a reasonable precision. The coherent backscattering effect \cite{wolf1985,vanalbada1985,tourin1997,larose2004a} (also referred to as weak localization) does give a direct estimation of the transport mean free path but our experimental configuration did not allow this special measurement since we could not place a laser sensor in the vicinity of the laser source. Yet we checked that the experimental intensity decay $I(t)$ gives an order of magnitude for $\ell^*$ that is consistent with the theoretical value.\\
 
\begin{figure}[htbp]
	\centering
		\includegraphics[width=8.6cm]{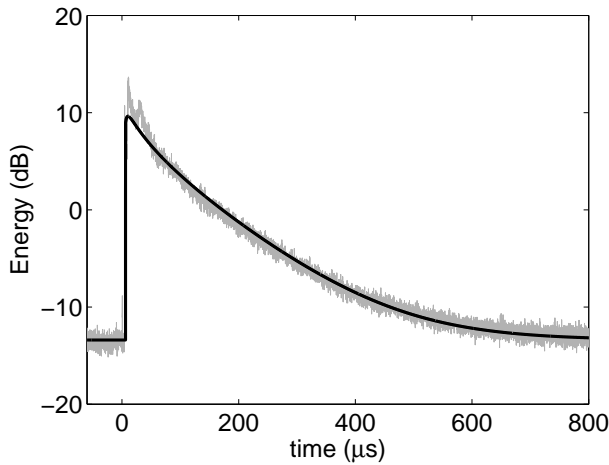}
	\caption{Averaged intensity $I(t)$ and theoretical fit for scalar wave diffusion in a 2-D semi-infinite slab with thickness $H=24$~mm. $\ell^*=5.5$~mm and $\tau_{abs}=120~\mu$s.}
	\label{fit_intensity}
\end{figure}

For the sake of simplicity we propose a 2-D scalar wave model for $I(t)$~\cite{tregoures2002c}, under the diffusion approximation.  In an infinite slab of thickness $H$ with perfect reflections on both sides, the averaged transmitted intensity reads:
\begin{widetext}
$$
I(X,t)=I_0\left\{\frac{1}{2H\sqrt{\pi Dt}} \quad+\quad\sum^{\infty}_{n=1} \frac{(-1)^n}{H\sqrt{\pi Dt}}e^{-\frac{n^2\pi^2 Dt}{H^2}}                     \right\}e^{-\frac{X^2}{4Dt}-\frac{t}{\tau_{abs}}}
$$
\end{widetext}
with $\tau_{abs}$ the absorption time (taking into account the intrinsic absorption in the aluminum and the lateral leaking due to the plasticine) and $D$ the diffusion constant. $X$ is the lateral distance between source and receiver. This formula is obtained using a modal decomposition of the diffusion equation in the $z$ direction. The intensity $I(t)$ is a mix of compressional and shear waves, each mode traveling with its own parameters (velocity, $\ell^*$, diffusion constant) and interchanging their energy through scattering events. In our experiment $\ell^*_S$ and $\ell^*_P$ are of the same order. The diffusion constant was approximated by $D=\frac{1}{2(1+2v_P^2/v_S^2)}(v_P\ell^*_P+2v_S(\frac{v_p}{v_s})^2\ell^*_S)\approx 40$~mm$^2$/$\mu$s. This assumption is valid after a couple of mean free times, when the equipartition regime~\cite{hennino2001} is set.  Equipartition means that the density of compressional and transverse modes equilibrates. Considering the specific velocities of each mode \cite{weaver1982}, we infer that 80\% of the energy is transported by S waves, and only 19\% by P waves (and an additional 1\% for surface waves). Hence our best fit (fig.~\ref{fit_intensity}) of the intensity decay in the coda gives $\tau_{abs}=120~\mu$s$ \pm 10\%$ and $\ell^*=5-20$~mm.

\section{Two-point correlation of diffuse fields}
In this section we focus on the experimental reconstruction of the direct Green's function from "passive" correlations. The main idea is to correlate diffuse fields sensed at two different locations on the top side when a source generates bulk waves at the bottom. Since we record the vertical component of the surface displacements, the two-point correlations should simulate a vertical source at the surface, which mainly generates surface waves. Indeed, the experimental correlations we obtained  reveal a wave packet that travels at the speed of a Rayleigh wave. We insist that our sources do not generate surface waves at the top side of the slab. Moreover if a surface wave happened to be generated anywhere, it would be completely absorbed by the plasticine. Under these conditions no Rayleigh wave should travel on the top surface, and no Rayleigh wave should be passively retrieved by correlations. Why then should passive imaging give rise to a Rayleigh wave train in our experiment?\\

We propose first to examine the role of scatterers for the emergence of the direct Rayleigh wavefront in the correlations. To that end we separately correlated coda records obtained through two different aluminum slabs: the first drilled with holes, the second without. Each impulse response was lasting $\approx 800~\mu$s before reaching the noise level (see fig.~\ref{signaux}). We underline that these record lengths are far from the Heisenberg time (break time) at which the modes of the aluminum block would be resolved (here $T_H \approx 10^6~\mu$s) and correlations would naturally converge to the Green's function. This modal approach is unrelevant to our experiment. The records were correlated and averaged over the 35 available sources. For the scattering slab, this reads:
$$
\left\langle C_{ij}(\tau)\right\rangle=\sum_{S=1}^{35}\int_{t=0~\mu s}^{t=600~\mu s}\varphi(S,X_i,t)\varphi(S,X_j,t+\tau)dt
$$
where $X_i$ and $X_j$ are the sensors points (running from $X_0=0$~mm to $X_6=60$~mm along the array). And for the homogeneous slab:

$$
\left\langle C^0_{ij}(\tau)\right\rangle=\sum_{S=1}^{35}\int_{t=0~\mu s}^{t=600~\mu s}\varphi_0(S,X_i,t)\varphi_0(S,X_j,t+\tau)dt
$$

\begin{figure}[htbp]
\includegraphics[width=8.6cm]{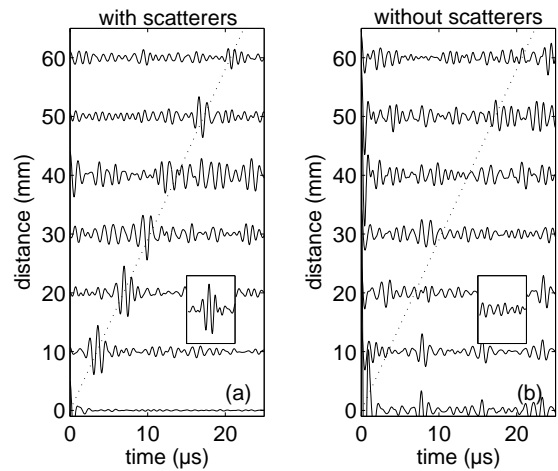}
\caption{\label{ResXCorr_F1} Green's function reconstruction for different pairs of receivers. Correlations are averaged over the 35 sources and filtered in the 0.8-1.6~MHz frequency band. On the left (a) are displayed the 7 cross-correlations $\left\langle C_{ij}(\tau)\right\rangle$ in the diffusive aluminum plate. On the right (b) the 7 cross-correlations $\left\langle C^0_{ij}(\tau)\right\rangle$ are obtained in an equivalent aluminum block without any hole, where the mode conversion possibly occurring  at the edges was suppressed by the plasticine. The insets show the summation of the 6 wavefronts from $X_0=10$~mm to $X_6=60$~mm after a time-reduction based on the Rayleigh wave speed  (dotted line, $v_R=2.9$~mm/$\mu$s).}
\end{figure}

\begin{figure}[htbp]
\includegraphics[width=8.6cm]{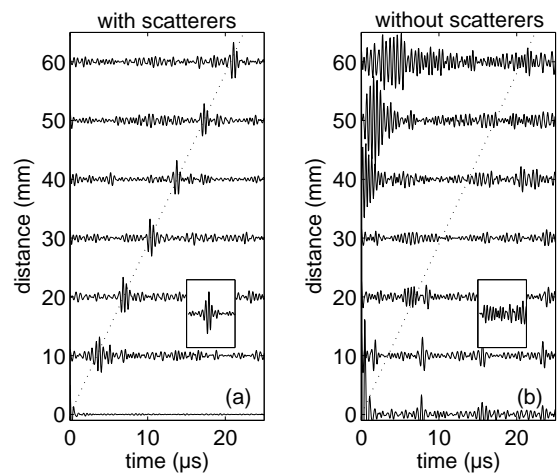}
\caption{\label{ResXCorr_F2} Same figure as fig.~\ref{ResXCorr_F1} except correlations are filtered in the 1.6-3.2~MHz frequency band. The signal-to-noise ratio in (a) is increased compared to the results obtained at lower frequencies.}
\end{figure}

To enhance the signal-to-noise ratio, each correlation is time symmetrized ($=\left\langle C(+\tau)\right\rangle+\left\langle C(-\tau)\right\rangle$) and normalized by its maximum. Results are displayed in fig.~\ref{ResXCorr_F1} and \ref{ResXCorr_F2}. A propagating wavefront (traveling at the Rayleigh wave velocity $v_R=2.9$~mm/$\mu$s) is clearly visible in the presence of scatterers, whereas it does not appear in the homogeneous slab. We also summed the 6 normalized propagating peaks after having delayed each signal according to the Rayleigh wave travel time. The summation is displayed in the enclosed box on each figure. In the scattering slab, its amplitude nearly corresponds to the coherent addition of 6 pulses. In the homogeneous device, the amplitude of the summation is $\approx 2.5$ (incoherent addition of 6 uncorrelated fields). We conclude on the necessity of mode coupling due to scattering for the Rayleigh wave to emerge from the passive correlations of diffuse fields generated by bulk waves sources. This is especially relevant for applications to seismology. Therefore, it appears once again~\cite{derode2003b} that the role of scattering is crucial in "passive imaging". Firstly, because of multiple scattering, at late times the equipartition regime can be attained whatever the sources/receivers positions. Secondly, because of mode conversions due to the scatterers, a Rayleigh wave emerges from the passive correlations even though no Rayleigh wave was generated by the sources. Note that the bandwidth in the upper band record is a little wider than in the lower band. This was done to compensate for the coda shortening ($<600~\mu$s) so that the product $T\Delta f$ is kept constant.\\

We can go a little further and catch a glimpse of the time symmetry properties. We performed the correlations into two consecutive time windows: from $0 $ to 45~$\mu$s and from 45~$\mu$s to 600~$\mu$s, and did not time-symmetrize the correlations (fig. 10 and 11). 45~$\mu$s is twice the time after which the diffuse energy spreads homogeneously along the array of receivers (length $L = 30$~mm) : $L^2/4D\approx 22~\mu$s in an open 2D scattering medium. The
time series were filtered in two frequency bands: 0.8-1.6~MHz and 1.6-3.2~MHz.\\

\begin{figure}[htbp]
\includegraphics[width=8.6cm]{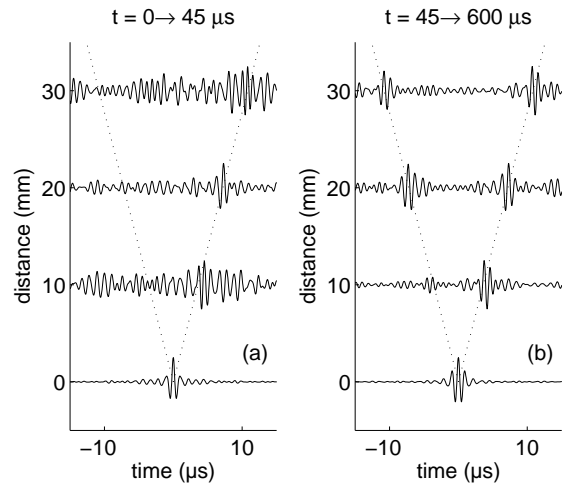}
\caption{\label{assymetryBF} Evolution of the asymmetry in the reconstructed Green's function in the low frequency regime (0.8-1.6~MHz), for early (a) and late (b) times. At early times, the waves are exciting scatterers on one side of the receiving array more than the other. The energy flux is clearly going from the source to the rest of the medium. After 45~$\mu$s the field is equipartitioned and the scattering halo fills the array, leading to a more symmetric correlation.}
\end{figure}

\begin{figure}[htbp]
\includegraphics[width=8.6cm]{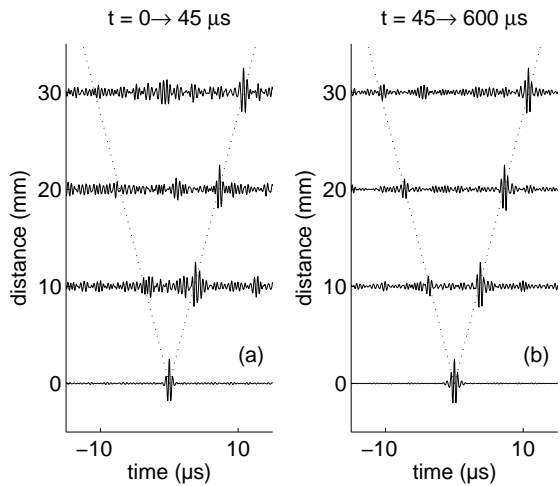}
\caption{\label{assymetryHF} Evolution of the asymmetry in the reconstructed Green's function in the high frequency regime (1.6-3.2~MHz), for early (a) and late (b) time windows. At early times, the waves are exciting scatterers preferably on one side of the receiving array. After 45~$\mu$s, bulk waves are expected to be equipartitioned, but not Rayleigh waves. Mode conversion to Rayleigh waves mainly occurs at one scattering cavity, leading to an anisotropic energy flux, and an asymmetric correlation.}
\end{figure}

In the first time-window, from 0 to 45~$\mu$s, correlations are asymmetric in time. This means that the \textsl{causal} part ($\tau>0$) and the \textsl{acausal} part ($\tau<0$) of the correlations are different (see left part of fig.~\ref{assymetryBF} and fig.~\ref{assymetryHF}). In the \textsl{causal} part, a  Rayleigh wavefront is clearly visible whereas noise is dominating the \textsl{acausal} part. This is due to the preferential direction of Rayleigh wave propagation (waves traveling from $X_0$ to $X_3$ in our experiment). There is a net flux of energy from $X_0$ to $X_1$, $X_2$ and $X_3$ (distances 10, 20 and 30~mm in fig.~\ref{assymetryBF} and fig.~\ref{assymetryHF}). This flux is due to the uneven distribution of sources in comparison to the receiver couples $X_0-X_1$, $X_0-X_2$ and $X_0-X_3$: most of the sources are on the $X_0$'s side. At the early times of the coda (from 0 to 45~$\mu$s), the diffusion regime is not yet attained.\\

 Later in the coda, from 45~$\mu$s to 600~$\mu$s, the wave field in the bulk of the aluminum slab is very likely to be equipartitioned. In the low frequency band (0.8-1.6~MHz), the time-symmetry of the correlation is indeed restored~\cite{vantiggelen2003,paul2005} (see right part of fig.~\ref{assymetryBF}): Rayleigh waves travel in all directions. Nevertheless and surprisingly, the asymmetry persists in the high frequency band (from 1.6~MHz to 3.2~MHz, see right part of fig.~\ref{assymetryHF}). To interpret this observation, we have carefully studied the location of the scattering sources around the array. In the high frequency band, the Rayleigh wavelength is $\sim 1.5$~mm. The generation of Rayleigh waves by scattering necessarily  occurs in the first half-wavelength beneath the free surface \cite{maeda2004}. In our scattering slab, one hole was nearly showing on the surface (position $X<0$), another one was $0.61$~mm beneath (position $X>60$~mm), the others being located much deeper. Rayleigh wave trains are mainly generated by the hole nearest to the surface, then propagate along the array of receivers (from $X_0$ to $X_6$ point). These waves are almost not perturbed (attenuated) until they reach the edges of the slab and the absorbing plasticine. They contribute  to a very clear propagating pulse in the positive part of the correlation. The weak coupling due to the deeper hole ($z=-0.61$~mm) on the $X_6$ side contributes to a smaller pulse propagating from $X_6$ to $X_0$ in the negative part of the correlations.\\

This interpretation is in agreement with observations in the low frequency band (0.8-1.6~MHz), where the average Rayleigh wavelength is 3~mm (fig.~\ref{assymetryBF}). At least 5 holes are present in the first half wavelength and should cause significant scattering of Rayleigh waves and coupling between surface and bulk waves. This time, the holes are evenly distributed along the sensor array. In the late coda, correlations $X_0\times X_1$, $X_0\times X_2$ and $X_0\times X_3$ are nearly symmetric. The time symmetry is obtained thanks to scattering by a symmetric distribution of scatterers. Under these conditions, a global equipartition among bulk and surface waves is guaranteed. Furthermore the reconstructed surface wave is strongly attenuated along its path because it senses many scatterrers along the array. In addition, these scatterers contribute signals around $\tau=0$ in the correlations, which degrade the reconstruction. We think this interpretation explains why the symmetric wavefront is much more noisy at low frequencies (fig.~\ref{ResXCorr_F1}), where lots of cavities are encountered in the first half wavelength, than at higher frequencies where the Rayleigh wavefront can propagate freely (fig.~\ref{assymetryBF}). \\

Finally, we comment on the possible misidentification of the waves that are reconstructed in our experiment.  Indeed,  if many scatterers are present within a wavelength, the overall wave
 velocity may be different (effective medium). In the heterogeneous plate,
a reduced-speed shear wave might propagate with the same wavespeed as a Rayleigh wave in the bulk aluminium plate (i.e. without cavities). In our experiment, the hole interspacing is 10~mm on average,  which is larger than the largest ultrasonic  wavelength. Thus, it is reasonable to assume that the shear waves propagate in the aluminium plate with the same wavespeed as in the bulk. The measured wave velocity is 2.9~mm/$\mu$s at all frequencies (from 0.8~MHz to 3.2~MHz) and indeed corresponds to the velocity of the  Rayleigh wave.\\

\section{Conclusion}
In this paper were presented laboratory experiments of elastic wave propagation in heterogeneous media at ultrasonic frequencies. An aluminum slab was made quasi-infinite by the use of absorbing boundaries to mimic the Earth's crust, in which scattering was obtained drilling cylindrical cavities. A relatively simple theoretical model for wave scattering properties was proposed. Wave field generation and detection was achieved using contactless and quasi-punctual laser devices. The cross-correlation of diffuse fields was performed, allowing us to retrieve passively the Rayleigh wave between two sensors only when scattering was present. Without scatterers, and in the case of bulk wave generation and surface detection, no Rayleigh wave was reconstructed. This illustrates the role of scattering and mode conversion in the Green's function passive reconstruction. \\

Analysis for different time-windows and frequency bands were conducted. Previous works in acoustics~\cite{weaver2005} and seismology~\cite{campillo2003a} observed that the Rayleigh wave reconstruction was harder with increasing frequency. Here we found that the Rayleigh wave reconstruction was more efficient in the high frequency band. In addition we observed that even in the late coda where waves are expected to be equipartitioned, asymmetry in the correlations may remain. Both observations are due to the very specific coupling between bulk and Rayleigh waves, which occurs if scatterers are present in the first wavelength beneath the free surface. We emphasize that equipartition of bulk waves does not always mean equipartition of surface waves. On the one hand, our experiments show the need of scattering to passively retrieve the impulse response between two sensors, on the other hand they show that scattering occurring between the sensors degrade the reconstructed Rayleigh waves. The trade-off between the two effects should be further investigated. Though our experiment was designed for seismological applications, results should be applicable to other fields of wave physics where both surface and bulk waves are present.

\appendix
\section{Calculation of the scattering cross-section of a cylindrical cavity.}
Here follows a brief calculation of the wave field scattered by a cylindrical cavity insonified by a  plane compressional wave. Three displacement potentials are relevant: $\Phi_i$ is the displacement potential of the incident P wave, $\Phi_s$ is the scattered P wave potential and $\Psi_s$ is the scattered S wave potential. They can be expanded as:

$$\Phi_i(r,t)=\sum_{m=0}^{\infty}\varepsilon^m i^m J_m(k_Pr)cos(m\theta)e^{-i\omega t}$$
$$\Phi_s(r,t)=\sum_{m=0}^{\infty}A_mH^{(1)}_m(k_Pr)cos(m\theta)e^{-i\omega t}$$
$$\Psi_s(r,t)=\sum_{m=1}^{\infty}B_mH^{(1)}_m(k_Sr) sin(m\theta)e^{-i\omega t}$$

where  $J_m$ and $H^{(1)}_m$ are respectively the Bessel and Hankel functions both of first kind and of order $m$ and $\varepsilon_m$ is the Neumann factor. Taking into account the null traction condition at the surface of the cylinder allows the calculation of the $A_m$ and $B_m$ coefficients. Those coefficients are given in \cite{paomow1973,faran1951,liu2000}. The scattering cross-sections are
 $$\sigma_{P\rightarrow P}= \frac{2}{k_P}[2|A_0|^2+ \sum_{m=1}^{\infty}|A_m|^2]$$
 $$\sigma_{P\rightarrow S}=\frac{2}{k_S}[\sum_{m=1}^{\infty}|B_m|^2]$$
 $$\sigma_P=\sigma_{P\rightarrow P}+\sigma_{P\rightarrow S}$$
  The corresponding differential scattering cross-sections are given by:

$$\frac{\partial \sigma_{P\rightarrow P}}{\partial\theta}(\theta)=\frac{4}{k_P} |\sum_{m=0}^{\infty}A_mi^{-m}cos(m\theta)|^2$$
$$\frac{\partial \sigma_{P\rightarrow S}}{\partial\theta}(\theta)=\frac{4}{k_S} |\sum_{m=1}^{\infty}B_mi^{-m}sin(m\theta)|^2$$

\begin{acknowledgments}
The authors wish to thank Richard Weaver and Julien de Rosny for fruitful discussions, Xavier Jacob and Samir Guerbaoui for experimental help. This work was supported by the Groupement de Recherche CNRS "Imagerie, Communication et D\'esordre" (GdR IMCODE 2253) and the CNRS program "DyETI".
\end{acknowledgments}

\newpage

\end{document}